\documentclass[oldversion]{aa}

\usepackage{graphicx}
\usepackage{epsfig}
\usepackage{color}
\usepackage{txfonts}
\usepackage{natbib}
\bibpunct{(}{)}{;}{a}{}{,} 

\begin{document}

\title{On the survival of metallicity gradients to major dry-mergers}
\titlerunning{Gradients and major dry-mergers}
\author{Paola Di Matteo$^{1}$, Antonio Pipino$^{2}$, Matthew D. Lehnert$^{1}$, Fran\c coise Combes$^{3}$, Benoit Semelin$^{3}$}
\authorrunning{Di Matteo et al.}
\institute{
$^{1}$  Observatoire de Paris, section de Meudon, GEPI,  5 Place Jules Jannsen, 92195, Meudon, France\\
$^2$ Department of Physics \& Astronomy, University of Southern California, Los Angeles 90089-0484, U.S.A.\\
$^3$ Observatoire de Paris, LERMA, 61 Avenue de l'Observatoire, 75014 Paris, France}

\date{Accepted,
      Received }


\abstract{Using a large suite of galaxies with a variety of concentrations
and metallicity gradients, we study the evolution of non-dissipative
(``dry'') equal mass mergers.  Our purpose in generating these simulations
is to explore how the metallicity gradient in dry mergers depends on
the structure and metallicity gradients of the galaxies involved in
the merger.  Specifically, we would like to answer: Could dry mergers
lead to metallicity gradients as observed in elliptical galaxies in the
local Universe? Do dry mergers always lead to a flattening of the initial
(i.e., pre-merger) gradient? From this modeling, we conclude that: The
ratio of the remnant and the initial galaxy slopes span a wide range
of values, up to values greater than 1 (with values greater than one
resulting only when companions have gradients twice the progenitor). For a
merger between two ellipticals having identical initial metallicity slopes
(i.e., equal companion and galaxy slopes), the metallicity profile of the
remnant flattens, with a final gradient about 0.6 times the initial one.
Ellipticals can maintain their original pre-merger metallicity gradient
if the companion slope is sufficiently steep.   The amount of flattening neither depends on the characteristics of the orbit of the progenitors or on their initial concentration. Given the diversity in outcomes of the mergers, we
conclude that dry mergers do not violate any observational constraints
on the systematic characteristics of metallicity gradients in local
ellipticals.  In fact, dry mergers may be important within the context
of the results of our simulations and may explain the large scatter in
gradients for massive ellipticals and the relative paucity of massive
ellipticals with no or shallow metallicity gradients.}


\maketitle

\section{Introduction}\label{sec:intro}

Recently, we have seen a revision in our understanding of the formation
of elliptical galaxies which was motivated by the need to reconcile
the apparently ``anti-hierarchical'' behavior of Active Galactic
Nuclei \citep[e.g.,][]{cat03, hasin05}, the evolution of their luminosity
function with redshift \citep[e.g.,][]{bundy06}, as well as the various
lines of arguments based on the analysis of their stellar populations
\citep{nelan05}. In particular, recent results suggest that we need a
substantial modification in how we treat the relationship between the
baryons and the dark matter in early type galaxies.  Pointedly, more
massive ellipticals are older and have formed faster than their lower
mass counterparts \citep{thomas05}.  In addition, independent evidence
that supports such \emph{downsizing} is inferred from the systematic
characteristics of abundance ratios in elliptical galaxies, namely, the
increase of the mean [Mg/Fe] in the stellar populations of ellipticals
with galaxy mass \citep{worthey92, matteucci94}.

Hierarchical modeling, in its most recent incarnations, only partly
accounts for the downsizing observed in elliptical galaxies. In
practice in such models, significant mass assembly still occurs at
late times, but most of the stars have been formed at high redshift in
small subunits that merge to grow early type galaxies.  The currently
preferred mechanism for the assembly of massive spheroids is a sequence
of ``dry'' mergers\footnote{In this paper \emph{dry merger} means a pure
dissipationless merger of an elliptical stellar systems, i.e., without any
gas or significant subsequent star formation.} \citep[e.g.,][]{khockfar03,
delucia06, cat06, cat08}.  There is evidence that dry mergers might solve some of the
outstanding issues related to the growth and evolution of elliptical
galaxies.  For instance, the brightest ellipticals have central
phase-space densities comparable to those of disk galaxies, suggesting
that, if mergers are responsible of their build-up, they likely require
only a small amount, if any, dissipation \citep{carlberg86}. Moreover,
dry mergers might explain the strong size evolution of massive red
galaxies  with redshift \citep[see, e.g., ][]{vandok08, khockfar06}
and the formation of slowly rotating ellipticals with boxy isophotes
\citep{naab06}.

One important aspect of determining whether or not dry mergers are a
viable mechanism for explaining the evolution of elliptical galaxies is
how various aspects concerning the metal content and distribution changes
after a merger event.  \citet{pipino08a} show that a series of multiple
dry mergers (with no associated star-formation) involving building-blocks
that satisfy the [Mg/Fe]-mass relation cannot fit the mass-metallicity
relation and vice-versa \citep[see also][]{ciotti91, cimatti06}.  A major
dry merger (mass ratios approximately about one), on the contrary, does
not violate these observational constraints if such a merger occurs
between galaxies that already obey both the mass($\sigma$)-[Mg/Fe] and
the mass($\sigma$)-metallicity relations. However, because this mechanism
would only operate over a limited range of initial masses, meaning,
you cannot have a long series of major mergers to make the most massive
ellipticals that initially obeyed the mass-metallicity relation, thus this
process alone cannot be the underlying physical cause of these trends.

But global trends and relationships amongst ellipticals is not the
only constraint available in testing the validity of the dry merger
scenario -- the structure within individual galaxies can provide further
significant constraints.  Radial negative metallicity gradients are a
common feature in the stellar populations of spheroids \citep{carollo93,
davies93}.  Observations show that the majority of ellipticals have
a typical decrease in metallicity of 0.3 dex per decade in radius
but with a large scatter \citep{annibali07}.  Moreover, a positive
correlation between the metallicity gradient with the galactic mass has
been originally claimed by \citet{carollo93}, but only for masses lower
than $\sim10^{11}M_{\odot}$.  More recently, additional studies have also
found a positive correlation of the gradient with mass for a wide range
of masses \citep[][but see \citealt{koba99, annibali07}]{ogando05, forbes05,sanchez07,spolaor09}.

Steep metallicity gradients are expected from classical dissipative
collapse models \citep[e.g.,][]{larson74, carlberg85}. More modern
versions of dissipative collapse starting from (semi-) cosmological
boundary/initial conditions \citep{kawata99, chiosi02} can also
explain the relation between metallicity gradient slope and stellar
mass.  The predicted metallicity gradient can be as high as $-$0.5 dex
decade$^{-1}$  in radius, hence galaxies can exhibit a slope
steeper than the average.  Owing to such a steep predicted slope it has
been hypothesised that galaxy mergers and monolithic collapse could be at
work together in order to explain the range in slopes \citep{koba04}.
On the other hand, more realistic models with a detailed treatment of the
chemical evolution predict a typical metallicity gradient of $-$0.25 dex
decade$^{-1}$ \citep{pipino08b} and also a trend between the gradient
and mass. These are just the trends necessary to explain the latest
observations \citep[][in preparation]{pipino09}.  This finding implies
that the assumption that ellipticals are formed through a mixture
several channels  \citep[including mergers and monolithic collapse,
e.g.][]{koba04} is not necessary, but cannot be excluded with  current available 
data.

N-body numerical simulations can be an important tool for understanding
the evolution of metallicity gradients in mergers and their remnants.
Such simulations are able to account for different environments, mass
ratios, and morphologies of the progenitor galaxies. The cosmological
simulations by  \citet{koba04} have shown that a variety of gradients in
early-type galaxies can be generated and are mainly due to a difference
in merging histories. In such a scenario, gradients are destroyed during
mergers, by an amount depending on the mass ratio of the progenitors.
Gradients can be regenerated if strong star formation occurs in the
central regions -- thus requiring sufficient gas dissipation to fuel
such star-formation -- and they slowly evolve through subsequent gas
accretion. These simulations indicate that, in order to reproduce
the large scatter observed in metallicity gradients, both gas infall
(monolithic collapse) and major mergers must occur.  While cosmological
simulations can follow a number of different merging histories (from
the coalescence of small gas-rich subunits to major dry or wet mergers),
it is also important to understand the role that each of these processes
play in reshaping any initial metallicity profile.

While a limited number of studies have investigated the evolution of
metallicity gradients in gas-rich mergers \citep{mihos94, bekki99},
taking into account both dissipative processes and the impact of star
formation, less attention has been devoted to the evolution of metallicity
profiles in dry mergers.  Early attempts at these types of simulations
suggested that metallicity gradients flatten, due to both mixing and
core-halo differentiation during the merger \citep{white80}. However,
N-body experiments have shown that the initial state of the galaxy is not
completely washed out during the coalescence of the systems \citep{van82},
suggesting that radial abundances may be only moderately reduced by
dissipationless mergers \citep{barnes96, mihos94}.

Given the importance of this problem for understanding the evolution
of early type galaxies, we wish to investigate impact of dry mergers
on galaxy evolution further through the use of N-body simulations of
equal-mass elliptical galaxies.  From this investigation, we want to
understand: 1) if dry mergers always cause the flattening of an initial
galaxy gradient; 2) under what conditions does this flattening occur; 3)
what is the magnitude of the flattening that is typically produced for
a range of initial conditions; and 4) how does this evolution depend
on orbital and morphological properties of the interacting systems.
To achieve this aim, we simulated a number of equal-mass dry mergers, with
a variety of orbital parameters, morphologies, and initial metallicity
gradients for each component of the merger.  We have chosen to begin investigating equal mass mergers, because they are likely to result in the greatest amount of mixing in phase-space and so, ultimately, they should cause the strongest variations in the initial (i.e. pre-merger) metallicity profiles.Ê

The main characteristic of
the models (initial conditions and numerical code used) are briefly
described in Sections~\ref{model} and \ref{code}, results are presented
and discussed in Section~\ref{results} and conclusions are drawn in
Section~\ref{concl}.

\begin{figure*}
 \begin{minipage}{0.2\textwidth}
   \centering
\includegraphics[width=3.6cm,angle=0]{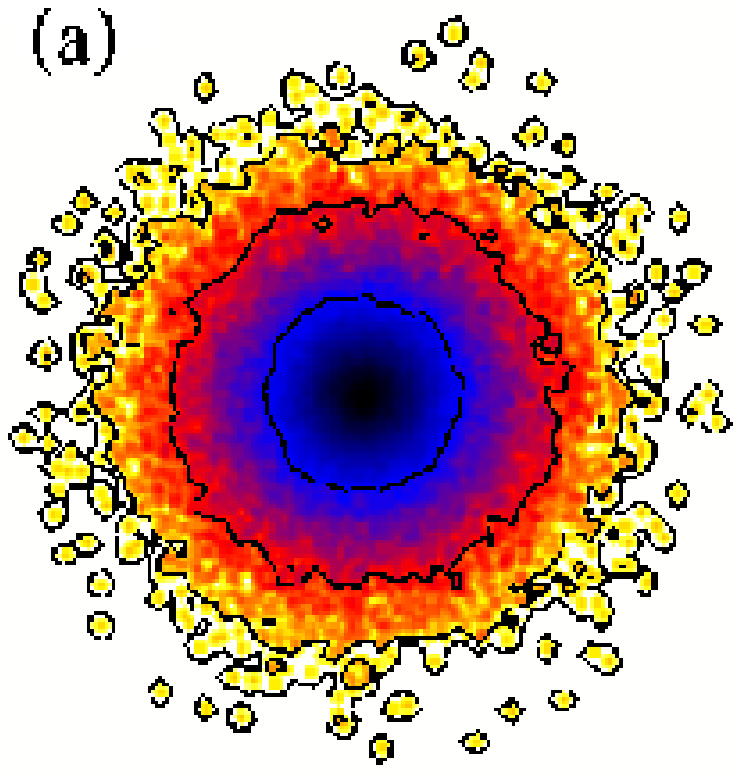}
 \end{minipage} \hspace{-0.2cm}
 \begin{minipage}{0.2\textwidth}
   \centering
\includegraphics[width=3.6cm,angle=0]{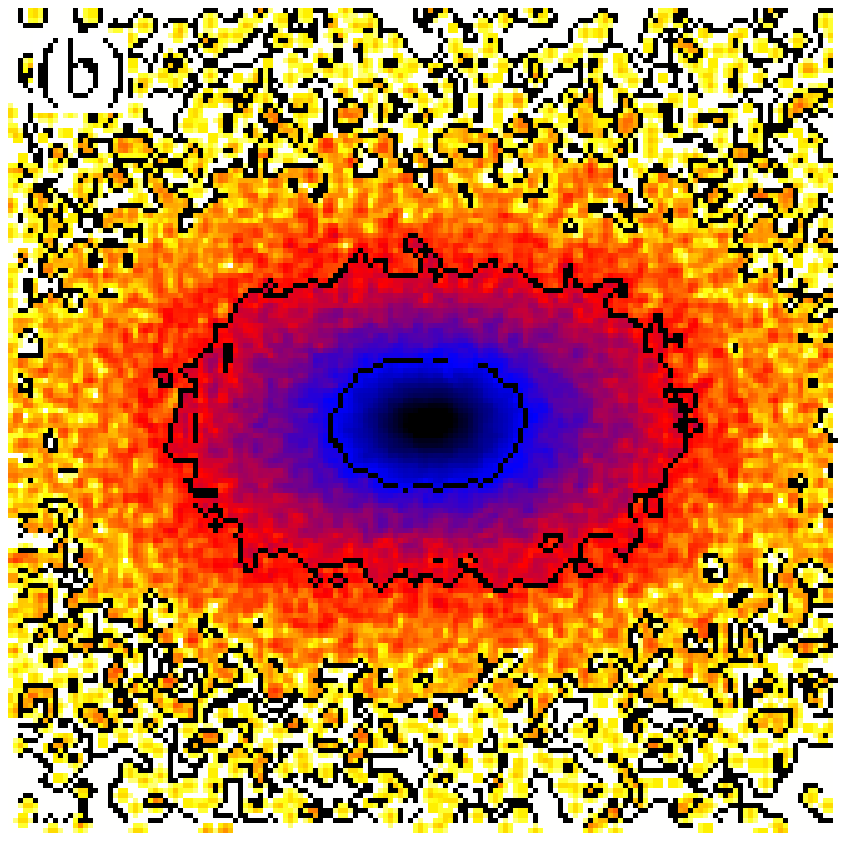}
 \end{minipage} \hspace{0.1cm}
 \begin{minipage}{0.1\textwidth}
   \centering
\includegraphics[width=4.1cm,angle=270]{fig3.ps}
 \end{minipage} \hspace{3.6cm}
 \begin{minipage}{0.1\textwidth}
   \centering
\includegraphics[width=5.5cm,angle=0]{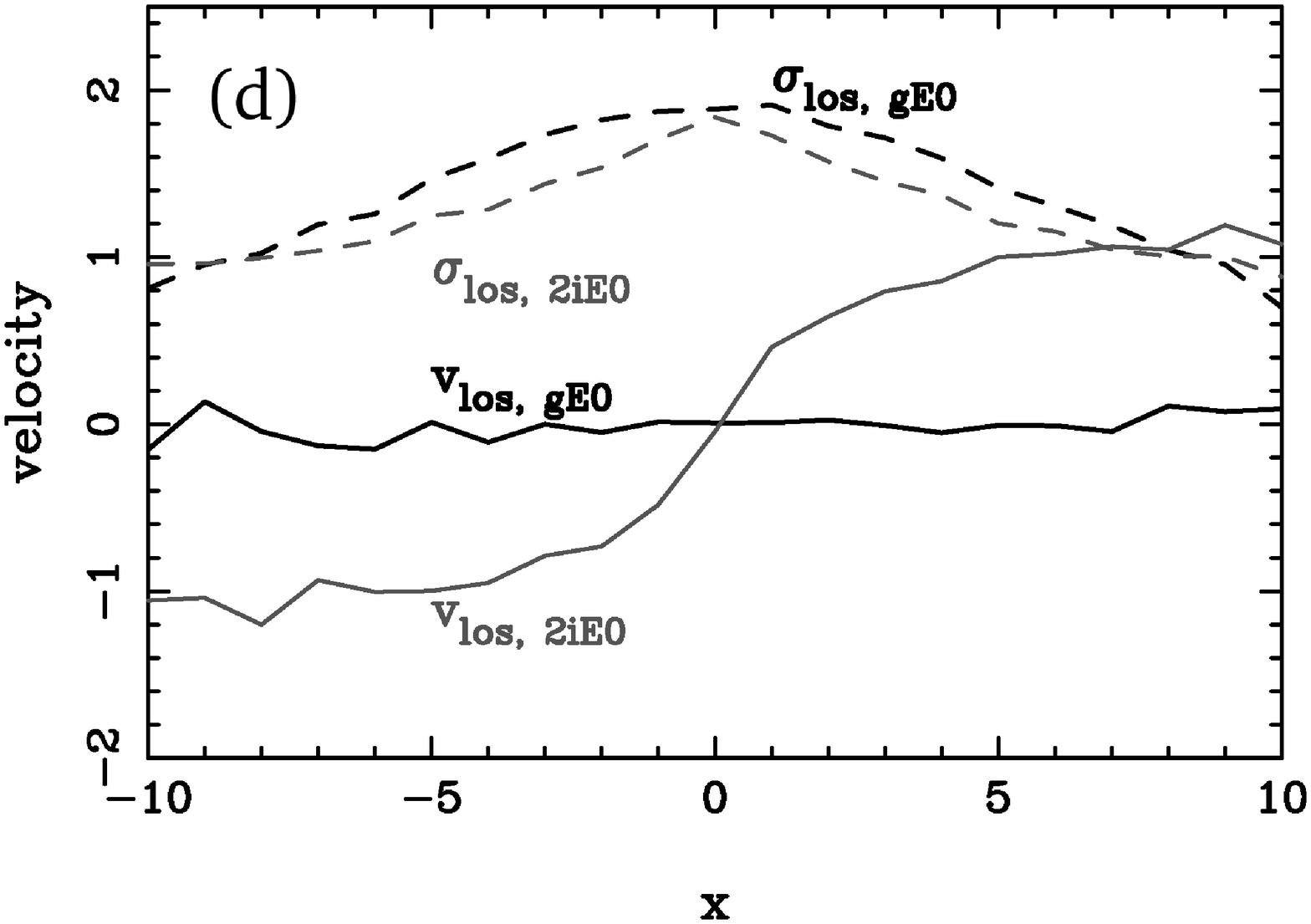}
 \end{minipage} 
\caption{Initial properties of some galaxy models: (a) projected-density
map of the gE0 galaxy, after evolving in isolation for 1 Gyr. Contours are equally spaced on a logarithmic scale;
(b) projected-density map of the 2iE0 galaxy which is the result of a
collision between two iE0s elliptical models (see text for details). We used the same scale interval and contours level as in the previous panel;
(c) volume-density profiles of some galaxy models, as indicated in the
legend of the figure; (d) line-of-sight velocity and velocity dispersion
profiles for the gE0 galaxy model (black solid and black dashed lines
respectively) and for the 2iE0 galaxy merger remnant (gray solid and
gray dashed lines respectively). \label{initfig}}
\end{figure*}

\section{The model}\label{model}

The mergers studied in this paper consist of the coalescence of two
equal-mass elliptical galaxies, without any dissipative component. The
fiducial case is described in Sect.~\ref{fiducial}. It corresponds
to mergers of two giant elliptical E0 (hereafter gE0) galaxies,
whose internal parameters are given in Table~\ref{galpar}, moving on 6
different orbits (2 parabolic, 1 hyperbolic, 1 radial and 2 ellipticals;
see Table~\ref{orbite}).

To understand to what extent the evolution of the metallicity gradients
depends on the galaxy parameters adopted, we also modeled interacting
galaxies having a central density $5$ times higher and $2.5$ times lower
than that of the gE0 galaxies  (we refer to these systems as gE0m and
gE0l, respectively; see Table~\ref{galpar} for their morphological
parameters). To allow the models to reach an equilibrium and to
demonstrate their initial stability, all the modeled galaxies were
allowed to evolve in isolation for 1 Gyr before any encounter. This
allows us to distinguish internal relaxation phenomena from those due
to the merging process itself.  These systems represent quite idealized
initial conditions: they are completely spherical, showing no rotation
in their baryonic component. Observations show, however, a vast range of
ellipticities and levels of rotation \citep[see, for example][]{emsell07}.
The systematics of the amount of rotation and ellipticity with mass
seem to suggest that more massive spheroids are not likely to be the
result of dissipative mergers \citep{carlberg86, vandok08}.  Therefore,
to test ``more realistic'' initial conditions, we modeled also mergers
between two equal-mass galaxies (hereafter called iE0), having masses equal to 0.5 times that of gE0s.
These galaxies were then subsequently merged to form a system (hereafter called 2iE0) whose total mass is equal to that of gE0s, but having initially a different morphology (triaxial instead of spherically symmetric) and a certain amount of rotation (see Table~\ref{galpar}
for the initial parameters of the iE0 galaxies and Fig.~\ref{initfig} for some properties of the initial galaxy models). In this way, we were able to, at least partially, account for more
complicated merger histories and hopefully more realistic merger
components in our simulations.

Each galaxy in the merging pair has been modeled with $N=120000$
particles, distributed among stars and dark matter. We also re-simulated
some cases, using a number of particles two and four times higher,
attempting to insure that the results presented here do not depend
on the numerical resolution adopted in our models.

Finally, for each of the galaxies involved in the interaction and
merger, we have adopted 15 different initial metallicity profiles.
These profile were constructed by assigning a metallicity, $Z$, to each
star particle in the system that depended exponentially on the distance
of the star from the galaxy center (the formal functions are given in
Table~\ref{gradienti}). This is also why the galaxies are allowed to
evolve in isolation for 1 Gyr.  This time allows us to ensure that the
metallicity gradient of the isolated galaxy is stable and remains as it was assigned. A large
variety of initial central metallicities and gradient slopes are thus
studied in this way, ranging from flat profiles (as it is the case of
gradient id=grad02 and grad05) to metallicity gradients around -0.3 dex
per decade in radius (grad01, grad11, grad12) to quite steep slopes
(gradient id=grad14, grad15). In this and in the next sections, the
metallicity gradient is defined as,

\begin{eqnarray}\label{formula}
\Delta&=&\frac{\Delta log(Z)}{\Delta log(r)}\\
&=&\frac{log(Z(0.1r_{50}))-log(Z(r_{50}))}{log(0.1r_{50})-log(r_{50})}\nonumber\\
      &=&log(Z(r_{50}))-log(z(0.1r_{50})),\nonumber
\end{eqnarray}

where $r_{50}$ is defined as the radius which contains half of the
baryonic mass (``baryonic half-mass radius'').  \\
In total, we analyzed 28 different simulations of major dry mergers, and, for each interaction, we studied 120 different combinations of the initial metallicity profiles, corresponding to 3360 different possibilities.

\begin{table}
\caption[]{Galaxy parameters. Both the stellar and dark mater profiles are
represented by Plummer models, having characteristic masses, respectively,
given by $M_{B}$ and $M_{H}$ and core radii given by $r_{B}$ and $r_{H}$}
\label{galpar}
 \centering
  \begin{tabular}{lcccc}
       \hline\hline
       &  gE0 & gE0l& gE0m& iE0\\
       \hline
       $M_{B}\ [2.3\times 10^9 M_{\odot}]$ & 70& 70& 70& 35\\
       $M_{H}\ [2.3\times 10^9 M_{\odot}]$ & 30& 30& 30& 15\\
       $r_{B}\ [\mathrm{kpc}]$ & 4.0&  6.0& 2.0& 2.8\\
       $r_{H}\ [\mathrm{kpc}]$ & 7.0& 7.0& 7.0& 5.0\\
       $N_{star}$ &80000 & 80000& 80000& 40000\\
       $N_{DM}$ &40000 & 40000& 40000& 20000\\
       \hline  
       \hline
  \end{tabular}
\end{table}

   \begin{table}
     \caption[]{Orbital parameters}
     \label{orbite}
     \centering
     \begin{tabular}{ccccc}
       \hline\hline
       orbit id & $r_{ini}$ &  $v_{ini}$ & $L^{\mathrm{a}}$  & $E^{\mathrm{b}}$ \\
         & [kpc] & [100 $kms^{-1}$] &  [$10^2kms^{-1}kpc$] &  [$10^4km^2s^{-2}$]\\
       \hline
       \\[0pt]
       01& 100.& 2.0&57.0&0.\\
       02& 100.& 3.0&59.0.&2.5\\
       05& 100.& 2.0&80.0&0.\\
       13& 100.& 0.0&0.0&-2.\\
       14& 70.0& 1.5&41.3&-0.66\\
       15& 70.0& 1.6&41.3&-1.57\\       
       \\[0pt]
       \hline  
       \hline
     \end{tabular}
\begin{list}{}{}
\item[$^{\mathrm{a}}$] It is the absolute value of the angular momentum of the unit mass, i.e., $L=\mid\bf{r} \times \bf{v}\mid $.
\item[$^{\mathrm{b}}$] It is the total energy of the relative motion,  i.e.,\\ $E={v}^2/2-G(m_1+m_2)/r$, with $m_1=m_2=2.3 \times10^{11}M_{\odot}$.
\end{list}
   \end{table}

   \begin{table}
     \caption[]{Initial gradients and metallicity profiles for the elliptical galaxies}
     \label{gradienti}
     \centering
     \begin{tabular}{ccl}
       \hline\hline
       Gradient id& $\Delta$ & Metallicity profile\\
       \hline
       \\[0pt]
       grad01& -0.26 & $z(r)=3z_{\odot}10^{-0.07r}$ \\
       grad02&  0.00 & $z(r)=z_{\odot}$  \\
       grad03& -0.13 & $z(r)=3z_{\odot}10^{-0.035r}$ \\
       grad04& -0.13 & $z(r)=1.5z_{\odot}10^{-0.035r}$ \\
       grad05&  0.00 & $z(r)=1.5z_{\odot}$ \\
       grad06& -0.22 & $z(r)=3z_{\odot}10^{-0.06r}$ \\
       grad07& -0.18 & $z(r)=3z_{\odot}10^{-0.05r}$ \\
       grad08& -0.15 & $z(r)=3z_{\odot}10^{-0.04r}$ \\
       grad09& -0.07 & $z(r)=3z_{\odot}10^{-0.02r}$ \\
       grad10& -0.04 & $z(r)=3z_{\odot}10^{-0.01r}$ \\
       grad11& -0.30 & $z(r)=3z_{\odot}10^{-0.08r}$ \\
       grad12& -0.33 & $z(r)=3z_{\odot}10^{-0.09r}$ \\
       grad13& -0.37 & $z(r)=3z_{\odot}10^{-0.10r}$ \\
       grad14& -0.41 & $z(r)=3z_{\odot}10^{-0.11r}$ \\
       grad15& -0.44 & $z(r)=3z_{\odot}10^{-0.12r}$ \\
       \\[0pt]
       \hline  
       \hline
     \end{tabular}
   \end{table}

\section{Numerical method}\label{code}

All the simulations have been run using the Tree-SPH code described
in \citet{benoit02}. A further description is given also in
\citet{dimatteo07} and in \citet{dimatteo08}. The code uses
a hierarchical tree method \citep{bh86} to evaluate gravitational
forces and a smoothed particle hydrodynamics method to simulate the
evolution of gas \citet {lucy77,gm82}. Since the work presented here
only investigates dry-mergers, only the part of the code evaluating
the gravitational forces acting on the systems has been used.

Gravitational forces are calculated using a tolerance parameter
$\theta=0.7$ and includes orders up to the quadrupole term in a multiple
expansion. A Plummer potential is used to soften the gravity at small
scales, with constant softening lengths  of $ \epsilon=280\ \mathrm{pc}$
for all particles. The equations of motion are integrated using a leapfrog
algorithm with a fixed time step of 0.5~Myr.

\begin{figure*}
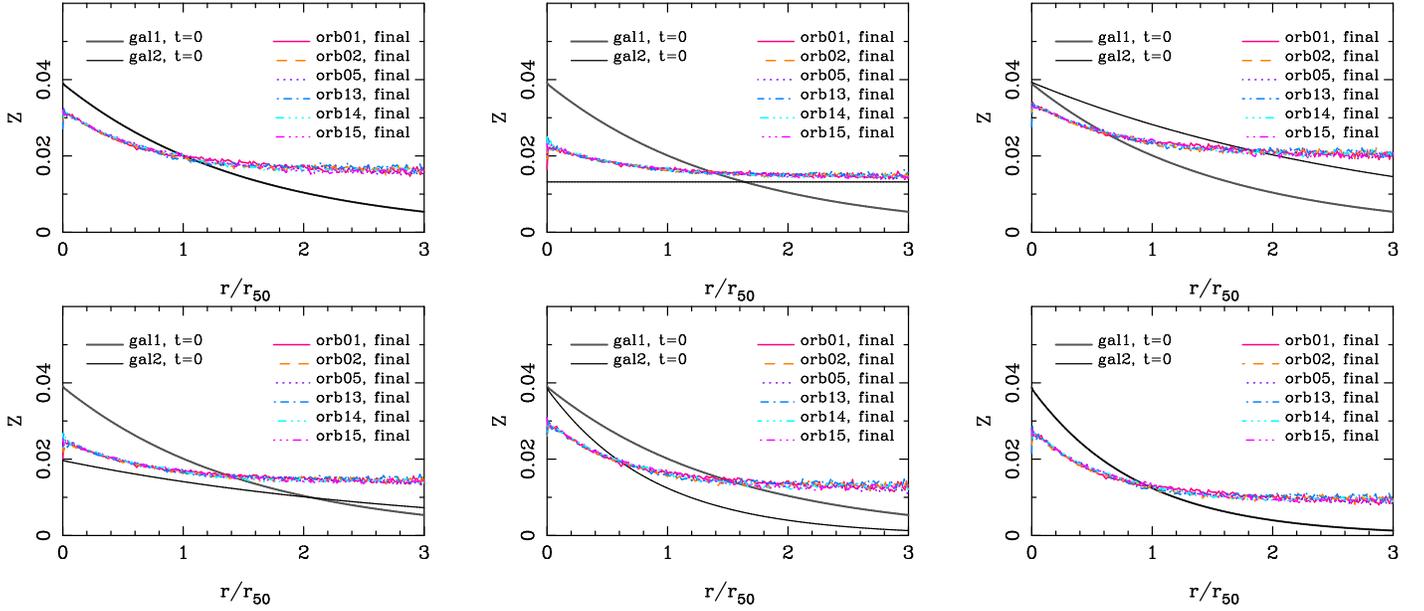

 \begin{minipage}{0.3\textwidth}
   \centering
\includegraphics[width=4cm,angle=270]{fig5.ps}
 \end{minipage} \hspace{0.8cm}
 \begin{minipage}{0.3\textwidth}
   \centering
\includegraphics[width=4cm,angle=270]{fig6.ps}
 \end{minipage} \hspace{0.8cm}
 \begin{minipage}{0.3\textwidth}
   \centering
\includegraphics[width=4cm,angle=270]{fig7.ps}
 \end{minipage}
 \begin{minipage}{0.3\textwidth}
   \centering
\includegraphics[width=4cm,angle=270]{fig8.ps}
 \end{minipage} \hspace{0.8cm}
 \begin{minipage}{0.3\textwidth}
   \centering
\includegraphics[width=4cm,angle=270]{fig9.ps}
 \end{minipage} \hspace{0.8cm}
 \begin{minipage}{0.3\textwidth}
   \centering
\includegraphics[width=4cm,angle=270]{fig10.ps}
 \end{minipage}
\caption{Evolution of the metallicity profiles in some major (1:1)
dry mergers of two gE0 galaxies. Metallicity is shown versus
the distance from the galaxy center for the two galaxies before
the interaction (black and thick gray solid lines) and for the
final remnant for different orbits (see figure legend). The values on
the abscissa are shown in units of the half-mass radius $r_{50}$ of
the corresponding galaxy. Different panels in the figure correspond to different
initial metallicity profiles for the two interacting galaxies:
{\it (top-left panel)} id=grad01grad01; {\it (top-middle panel)} id=grad01grad02; {\it (top-right panel)} id=grad01grad03; {\it (bottom-left panel)} id=grad01grad04; {\it (bottom-middle panel)} id=grad01grad15; {\it (bottom-right
panel)} id=grad15grad15.
\label{zprofile}}
\end{figure*}

\section{Results and discussion}
\label{results}

\subsection{Evolution of metallicity profiles in major dry mergers}
\label{fiducial}

The evolution of some of the metallicity profiles during major
non-dissipative mergers are shown in Fig.\ref{zprofile}. In this plot
the initial morphologies of the two interacting galaxies are always
the same, while the orbits and initial metallicity gradients have been
varied. In some cases, as for the top-left and bottom-right panels,
the two interacting galaxies have initially the same profile. This
situation of having identical initial equal metallicity profiles is the
case studied by \citet{white80}: in this cases, the profile flattens,
due to the redistribution of stars in the systems induced by the merging
process. Indeed, one can clearly see from the final profile that the
flattening is due to both the decrease of the central metallicity
and to the contemporaneous increase of the metallicity in the outer
regions. Note also that the evolution of the profile does not seem to be
very sensitive to the orbital initial conditions: orbits with different
initial energies and angular momenta produce very similar final profiles,
at least in the region inside the remnant half-mass radius. This happens because mixing and phase-space density evolution are independent on the orbital parameters, as discussed in Appendix \ref{app1}.\\

\begin{figure}
   \centering
\includegraphics[width=5cm,angle=270]{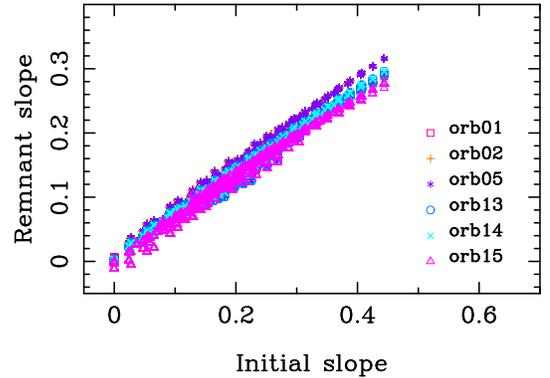}
\caption{Absolute values of the initial and final metallicity slopes for gE0-gE0 mergers. Different orbits are indicated by different symbols, as explained in the legend. The initial slope has been evaluated by  superposing  the initial metallicity profiles of the two progenitor galaxies. \label{superpos}}
\end{figure}

\begin{figure*}
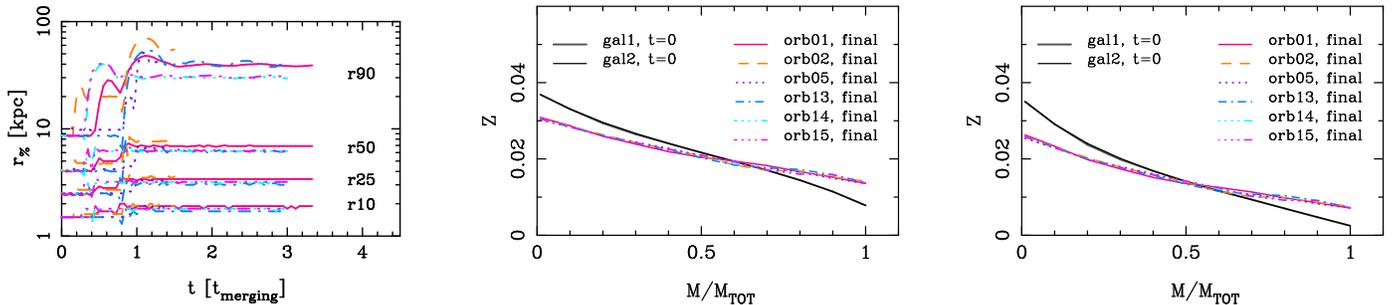

 \begin{minipage}{0.3\textwidth}
   \centering
\includegraphics[width=4cm,angle=270]{fig12.ps}
 \end{minipage} \hspace{0.8cm}
 \begin{minipage}{0.3\textwidth}
   \centering
\includegraphics[width=4cm,angle=270]{fig13.ps}
 \end{minipage} \hspace{0.8cm}
 \begin{minipage}{0.3\textwidth}
   \centering
\includegraphics[width=4cm,angle=270]{fig14.ps}
 \end{minipage}
\caption{Core-halo differentiation and mixing in some of major gE0-gE0
mergers modeled for this study. ({\it{left}}) Shows the evolution with
time of the Lagrangian radii containing 10$\%$, 25$\%$, 50$\%$, 75$\%$,
and 90$\%$  of the total stellar mass. For simplicity, the radii are
shown only for stars belonging to one of the two galaxies (the behavior
is the same for the companion). The time axis is in units relative to
the merging time. Mergers with a variety of orbital characteristics are
 shown, as indicated into the figure legend. ({\it{Middle and right}})
Metallicity profiles for some gE0-gE0 mergers, as a function of
the fractional mass contained.  The panels correspond to different
metallicity profiles: id=grad01grad01 {\it (middle panel)}; id=grad15grad15
{\it (right panel)}. Metallicity is shown for the two galaxies before the
interaction (black and thick gray solid lines) and for the final remnant.
\label{chvsmix}}
\end{figure*}

Fig.~\ref{zprofile} also shows some examples of the two galaxies
having different initial metallicity gradients and distributions
(``mixed''). As discussed in the Introduction, observations \citep[see,
for example][]{ogando05,spolaor09} show that early-type galaxies in the local
Universe have a wide dispersion in their metallicity gradients,
particularly at the high end of the distribution of masses. This means
that dry mergers between galaxies with different metallicity gradients
should be common locally. Of course, this is not true if there is some
sort of regularity in or correlation between the metallicity gradients
as a function of local galaxy density, for which yet there is no clear
evidence \citep[see][]{sanchez06, clemens08}.  But if ``mixed'' mergers
are common (as seems likely), then it is interesting to understand
what the resulting gradient is and how it depends on the initial
profiles and orbital parameters of the progenitors. By examining the
plots (Fig.~\ref{zprofile}) and analyzing the simulations, the resulting
gradients in mixed mergers can lie in between those of the progenitors:
the merging of galaxies with a flat and steep gradient ($\Delta=-0.26$)
produce a remnant having a slope of about -0.09 which is lower than the
steepest progenitor gradient but certainly not flat as in the companion
galaxy.  This implies that for a given progenitor galaxy with an initial
metallicity gradient,  \emph{major dry mergers do not necessarily lead
to a flattening of the initial galaxy slope, but  the outcome depends greatly on the
slope of the companion galaxy.} An early-type elliptical galaxy merging
with an other having the same initial metallicity profile will have
a final slope lower than the initial one -- the amount of flattening
is evaluated in Sect.\ref{amount}-- but if it merges with a companion
having a sufficiently steep profile, the final slope can be the same
or steeper than the initial one. \\
Comparing the superposition of the initial metallicity profiles of the two progenitor galaxies (i.e. $Z_1(r)+Z_2(r)$) with the remnant profile, we find that, on average, the final metallicity gradient is a factor 0.6-0.7  flatter than the initial one, with only a small dispersion around the average (see Fig.\ref{superpos}).

\subsection{Core-halo differentiation and phase-space mixing}
\label{mixing}

In Sect.\ref{fiducial}, we showed that the metallicity gradient of the
final merger remnant depends on the initial gradients of the progenitor
galaxies. Here we present more details about the mechanisms causing
the evolution of these profiles. Importantly, we note that the metallicity
profiles given in Fig.\ref{zprofile} are shown in units of the stellar
half mass radii $r_{50}$ of the progenitors and remnant. Thus,
a variation in the half-mass radii of the remnant with respect to that of
the progenitor galaxies, without any variation in the Z(r) profile, should
be sufficient, in principle, to change the final metallicity profile.

Indeed, as shown for gE0-gE0 mergers in the left panel
of Fig.\ref{chvsmix}, the half-mass radius changes with time, at each pericenter passage and 
mostly during the final phases of coalescence of the two systems
(t/t$_{merging}\simeq$1), with the final value being about 50$\%$ greater
than the pre-interaction value. Note also that outer parts of the systems
(those outside $r_{50}$) show the strongest evolution. Indeed, the tidal
shocks increase  the radius $r_{90}$, containing  90$\%$ of the stellar
mass by more than a factor of 3. On the contrary, the inner radii show
only modest change after the merger.  Moreover, the evolution with radius
also has a behavior that depends on the orbital geometry of
the progenitors.  The dynamical evolution of the outer radii is affected by the
orbital geometry while that of the inner regions is largely independent
of the orbital parameters.

As first shown by \citet{white80}, it is possible to separate the effects
due to structural changes from those arising from mixing by showing
the metallicity profiles as a function of the enclosed mass rather
than as a function of some physical coordinates.
If the evolution of the metallicity gradient was only due to core-halo
differentiation, then the initial and final profiles in enclosed mass
would be identical.  However, a clear flattening occurs during the
merger (see  Fig.~\ref{chvsmix}, middle and right panels), indicating that mixing must contribute to the evolution of
the metallicity gradient. In our simulations, this flattening is more
pronounced than that found by \citet{white80}. This difference could
be due to his choice of concentrating on merger remnants which showed
density profiles and velocity structures as close as possible to that of
the progenitors \citep[see discussion in ][]{white80}.  In addition,
the process of mixing does not seem to depend on the initial orbital
geometry but it is generic to the merging process itself and leads to
rather similar metallicity gradients.  This implies that mixing is
mostly driven by the progenitor structure that (at least partially)
survives the merging process \citep[see also][]{white80, barnes96,
valluri07}. As a consequence, our simulations indicate that it is not
possible to disentangle the orbits of the parent galaxies by using the
metallicity distribution of the remnant if the analysis is restricted
to the central region of the galaxy (i.e., inside the half-mass radius).

\begin{figure}
   \centering
\includegraphics[width=5cm,angle=270]{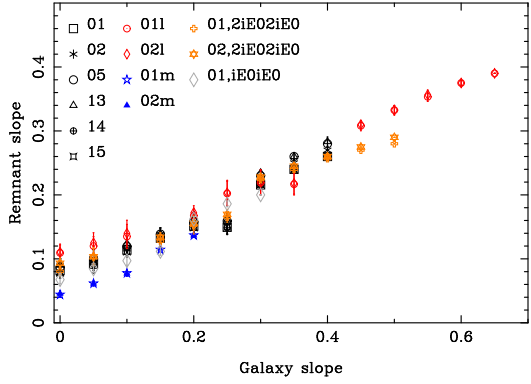}
\caption{Remnant (absolute) slopes as a function of the initial galaxy
(absolute) slopes, for different orbits and different initial galaxy
models. Symbols '01', '02', '05', '13', '14' and '15' in the legend
correspond to interactions among two gE0 galaxies moving on different
orbits (see Table~\ref{orbite} for the corresponding orbital parameters);
'01l' and '02l' correspond to interactions among two gE0l galaxies
with orbit id=01 and 02, respectively; '01m' and '02m' correspond
to interactions between two gE0m galaxies  with orbit id=01 and 02,
respectively. Elliptical galaxies resulting from the merger of two 2iE0
galaxies are shown (with the symbols '01,2iE02iE0' and '01,2iE02iE0'
referring to orbit id=01 and 02, respectively). For comparison, also the
slope resulting from the merger of two iE0 galaxies -- having a total mass
half of that of gE0 -- are shown. Error bars correspond to the standard
deviation of the mean. \label{gradients}}
\end{figure}

\begin{figure}
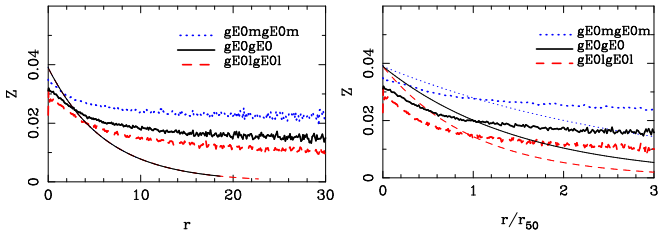

   \centering
\includegraphics[width=3cm,angle=270]{fig16.ps}
\includegraphics[width=3cm,angle=270]{fig17.ps}
\caption{Left panel: Evolution of the metallicity profiles for three different major dry mergers, involving respectively two gE0 galaxies (black solid line), two gE0l galaxies (dashed red line) and two gE0m galaxies (dotted blue line). The initial profile is shown with corresponding thin curves. All the profiles are shown as a function of the distance r form the galaxy center.  Right panel: Same as the previous panel, but this time all the profiles are shown in units of the half mass radius $r_{50}$ of the corresponding galaxy.  \label{denscomp}}
\end{figure}

\begin{figure}
\centering
\includegraphics[width=5cm,angle=270]{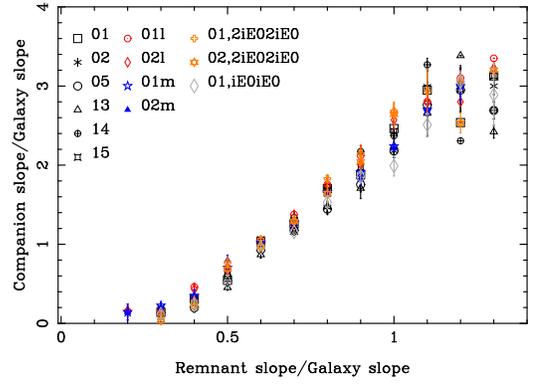}
\caption{Ratio of the companion (initial) slope to the galaxy (initial)
slope versus the ratio of the remnant slope to the galaxy (initial)
one. Symbols are described in caption of Fig.\ref{gradients}. Error bars
correspond to the standard deviation of the mean. Note that the galaxies having a remnant slope of zero are not plotted in this figure.  This corresponds
to the situtation where the initial galaxy slope and companion slope are
both zero and thus 1/(galaxy slope) is infinite.
\label{ratio}}
\end{figure}

\subsection{Metallicity gradients in major dry mergers: flattening or
steepening?}\label{amount}

After describing how metallicity profiles evolve during a merger and
the physical processes which determine this evolution, we now want to
quantify the change in the radial profiles themselves as
a function of the initial metallicity distribution. 

In Table \ref{finslope_1to1}, we tabulate the initial and final metallicity
gradients for simulations of two gE0 galaxies with a specific orbital
geometry (id=01) but for a range of metallicity gradients. To generalize
this, Fig.\ref{gradients} shows the absolute values of the initial and
final gradients from the simulations for all the galaxy morphologies
and orbits.  For a given initial metallicity gradient there are a
range of final gradients, depending mainly on the profile of the companion
galaxy. The initial galaxy models span a wide range in metallicity slopes
,0.0- -0.7, while the remnant slope only spans about 0.0- -0.4.  
 Also, a close inspection of Table~\ref{finslope_1to1} tells us that the final slope is always flatter than
the maximum between the parent and his companion slopes. In most of the cases, 
the final slope is lower than the mimimum between the parent and his companion slopes.
Note that,
as suggested previously, galaxies having the same morphologies have very
similar final slopes, independent of their orbital geometry. This is the
case regardless of the initial morphologies.  In addition, the most concentrated
galaxies (gE0m) have the smallest range of gradients and generally
smaller gradients for the same initial conditions while galaxies with
low concentrations (e.g., gE0l) generally have larger slopes and exhibit
a wider range of slopes. This is due to the fact that we have assigned
to each galaxy the metallicity profiles following a simple Z=Z(r) law
where r is the physical distance from the galaxy center of mass. That is,
the initial metallicity profiles have not been normalized in anyway to
the galaxy half-mass radii, while the metallicity gradients, according
to Eq.\ref{formula}, are evaluated using them. \\A comparison of the evolution of the metallicity profiles for galaxies with different initial concentration is shown in Fig.\ref{denscomp}. In the left panel,  the initial metallicity profile is shown as a function of r, the distance from the center, for the gE0l, gE0 and gE0m galaxy. This initial profile is obviously the same for the three galaxies, accordingly to our choice. The final profile of the remnant galaxy, in turn, depends on the initial galaxy density, in the sense that the more concentrated the galaxy is, the higher is the value of Z(r), for any given r. This is simply a consequence of the fact that the more concentrated the galaxy is, the more metal rich it is too, according to the way the initial metallicity profile has been assigned. When normalizing to the half-mass radius (right panel in Fig.\ref{denscomp}), in turn, the initial profiles are obviously different, as well as the final ones, but interestingly the amount of flattening of the final gradients with respect to the initial ones are very similar. If the two galaxies have initially identical metallicity profiles, as in the case shown in Fig.\ref{denscomp}, the final metallicity gradient will have a slope equal to 0.6-0.7 times the initial one, with no clear correlation with the initial central density of the galaxy.\\

Fig.\ref{gradients} shows that a range of initial gradients gives rise to
a variety of remnant slopes, this depending, of course, on the slope of
the companion galaxy, as well as on the initial concentration.  What is
the relationship between the remnant and initial galaxy slopes?  This is
shown in Fig.\ref{ratio}.  The relation between the initial and remnant
gradients has generally a small scatter about a mean relation, with
higher dispersion for the high ratios. This figure basically summarize
the results of this work which implies:

\begin{itemize}

\item The ratio of the remnant and the initial galaxy slopes span a wide
range of values, up to values greater than 1.

\item For a merger between two ellipticals having identical initial
metallicity slope (i.e., companion slope=galaxy slope), the metallicity
profile of the remnant flattens, with a final gradient about 0.6 times the
initial one.

\item Remnants can have metallicity gradients greater than that of the
progenitor elliptical galaxy. Our simulations show that this happens
every time the companion has a slope two times greater than the parent.

\item Ellipticals can maintain their original pre-merger metallicity
gradient if the companion slope is sufficiently steep.

\item The final remnant gradient does not depend on the orbits of the
progenitor ellipticals.

\item While the final gradient does depend on the initial concentration of the progenitors, the amount of flattening (i.e. remnant slope/progenitor slope) does not.

\end{itemize}

\subsection{The Role of Dissipationless Mergers in the Evolution of
Elliptical Galaxies}

\begin{figure}
\includegraphics[width=6cm,angle=270]{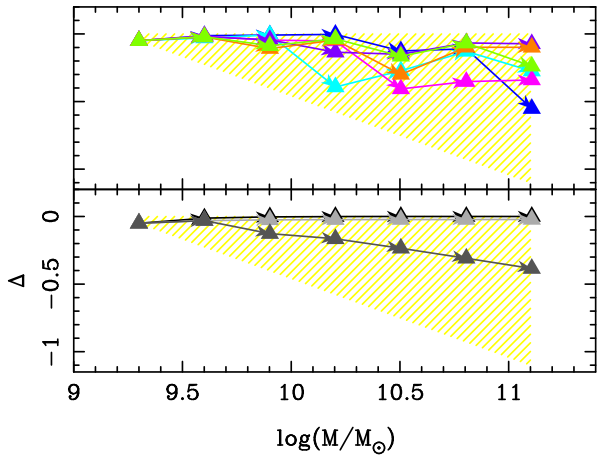}
\caption{Evolution of metallicity gradients through a sequence of major
dry mergers. The initial elliptical galaxy has log(M/M$_\odot$)=9.3
and $\Delta$=-0.05. ({\it{Upper panel}}) Different colors correspond
to different merging histories, with companion gradients randomly
distributed in the shaded area (cf. Fig.2, right panel, in Ogando et
al. 2005). ({\it{Lower panel}}) Different colors correspond to different
merging histories, with companion gradients always equal to zero (black
triangles), to the initial galaxy gradient (light gray triangles), and
to the maximum slope allowed for a given mass, i.e., lower edge in the
shaded area (dark gray triangles). \label{chemin}}
\end{figure}

Do these results in any way suggest that equal mass dry mergers might be
inconsistent with the metallicity gradients observed in local ellipticals?
The short answer is no.  There are no systematic trends in magnitude
of metallicity gradients in ellipticals save one, metallicity gradients
generally become steeper with increasing mass, but with a large scatter
\citep{ogando05,spolaor09}.  Our models show that, as long as the metallicity
gradient of the companion is not more than twice that of the parent,
the remnant will have a shallower gradient.  Thus one can imagine that
the build up of the most massive ellipticals is through a series of dry
mergers whereby the gradient gradually softens leading to a trend, albeit
with large scatter, between mass and metallicity gradient.  But  if
there are several mergers in sequence with long times between then,
the metallicity gradients would increase and decrease rather randomly,
thus leading to a situation where there are few systematic trends
between the characteristics of the metallicity gradient and structure
properties of ellipticals. This is precisely shown in Fig.\ref{chemin},
where the evolution of an initial metallicity gradient, due to recursive
dry mergers, is followed. If dry mergers can occur among ellipticals
having different gradients,  randomly distributed in the shaded
area\footnote{This area encloses the negative metallicity slopes found by
Ogando et  al. (2005), and as shown in their Fig.2.} shown in
the plot, a sequence of such events does not lead to a clear trend (for
example, the more massive is the galaxy, the shallower its profile). In
turn, \emph{a sort of random walk, with increasing and decreasing slopes,
seems to be the general outcome} (upper panel in Fig.\ref{chemin}), unless at
each merger epoch, a given galaxy is allowed to merge only with companions
having the identical gradient (lower panel in Fig.\ref{chemin}). 
Because of the stochastic nature of the process of dry mergers, it is
difficult to gauge the frequency, and therefore the importance of this
process in galaxies.  While, most likely, the merger of two similar
ellipticals would lead to a lessening of the metallicity gradient, the
final product depends on the concentration of the individual galaxies
and their initial gradients.  As the gradient of the parent galaxy
because flatter and flatter, it then becomes likely that another dry
merger will actually increase its metallicity gradient.  If some of the
mergers are not completely dissipationless, then this will also tend to
increase the gradient.  However, the relative mass of gas that forms stars
during any gas accretion or merger must be small. If the fraction of mass
forming stars were large or did so over extended periods of time, then
this process would violate the well-known enhancement in $[\alpha/Fe]$ in
ellipticals, especially the most massive ones.  Therefore, any enhancement
in the gradient due to star-formation must be small and the influence
of the dissipationless component must be dominant in determining the
final metallicity gradients in massive elliptical galaxies.

One of the obvious outcomes of all of this is that if ellipticals at all
masses were formed with the same gradient then we would expect the scatter
in the metallicity gradient as a function of mass to increase with both
time and mass.  This is simply because our simulations show that
metallicity gradients that are initially steep will most likely soften.
This will continue up to the point where the gradient is shallow enough
to make it more likely that the merger will be with a galaxy that has a
slope steep enough (by a factor of 2) to actually increase the gradient in
the remnant.  This variation in outcomes would thereby have the effect of
increasing the scatter as galaxies grow more massive.  This is similar to
what has been observed \citep[][but see \citet{annibali07}]{ogando05}.
It may also explain the relative lack of galaxies with no gradients
among the massive ellipticals \citep[e.g.,][]{ogando05, annibali07}.
If the gradients became flat, then they are likely to merge with galaxies
that are steep enough to increase their slopes.

Note that, in this analysis,
the metallicity gradients have been chosen to reflect
those of elliptical galaxies in the local Universe. The progenitors
of the most massive ellipticals observed today probably
were low and intermediate mass ellipticals that merged at early epochs,
and possibly had structural properties different from those of local
galaxies. Unfortunately, the properties of distant early type galaxies are not
well determined.  This leaves no recourse but to use the observed
properties of local galaxies.   Thus we caution that our analysis
may not capture in detail the cosmological evolution of elliptical
galaxies.

   \begin{table*}
      \caption[]{Initial and final metallicity gradients for the fiducial 1:1 dry mergers.}
         \label{finslope_1to1}
     \centering
      \begin{tabular}{cccccccc}
        \hline\hline
        Run Type& Initial slope & Initial slope & Final slope & Run Type& Initial slope & Initial slope & Final slope\\
                & (Galaxy 1) & (Galaxy 2) & (Remnant) & & (Galaxy 1) & (Galaxy 2) & (Remnant) \\
        \hline

grad01grad01 &-0.26 &-0.26 &-0.17  & grad05grad11 &0.00 &-0.30 &-0.11  \\
grad01grad02 &-0.26 &0.00 &-0.09  & grad05grad12 &0.00 &-0.33 &-0.12  \\
grad01grad03 &-0.26 &-0.13 &-0.13  & grad05grad13 &0.00 &-0.37 &-0.13  \\
grad01grad04 &-0.26 &-0.13 &-0.13  & grad05grad14 &0.00 &-0.41 &-0.13  \\
grad01grad05 &-0.26 &0.00 &-0.08  & grad05grad15 &0.00 &-0.44 &-0.14  \\
grad01grad06 &-0.26 &-0.22 &-0.16  & grad06grad06 &-0.22 &-0.22 &-0.15  \\
grad01grad07 &-0.26 &-0.18 &-0.15  & grad06grad07 &-0.22 &-0.18 &-0.14  \\
grad01grad08 &-0.26 &-0.15 &-0.14  & grad06grad08 &-0.22 &-0.15 &-0.13  \\
grad01grad09 &-0.26 &-0.07 &-0.11  & grad06grad09 &-0.22 &-0.07 &-0.10  \\
grad01grad10 &-0.26 &-0.04 &-0.09  & grad06grad10 &-0.22 &-0.04 &-0.08  \\
grad01grad11 &-0.26 &-0.30 &-0.18  & grad06grad11 &-0.22 &-0.30 &-0.17  \\
grad01grad12 &-0.26 &-0.33 &-0.19  & grad06grad12 &-0.22 &-0.33 &-0.18  \\
grad01grad13 &-0.26 &-0.37 &-0.20  & grad06grad13 &-0.22 &-0.37 &-0.18  \\
grad01grad14 &-0.26 &-0.41 &-0.20  & grad06grad14 &-0.22 &-0.41 &-0.19  \\
grad01grad15 &-0.26 &-0.44 &-0.21  & grad06grad15 &-0.22 &-0.44 &-0.19  \\
grad02grad02 &0.00 &0.00 &0.00  & grad07grad07 &-0.19 &-0.18 &-0.13  \\
grad02grad03 &0.00 &-0.13 &-0.08  & grad07grad08 &-0.19 &-0.15 &-0.11  \\
grad02grad04 &0.00 &-0.13 &-0.05  & grad07grad09 &-0.19 &-0.07 &-0.09  \\
grad02grad05 &0.00 &0.00 &0.00  & grad07grad10 &-0.19 &-0.04 &-0.07  \\
grad02grad06 &0.00 &-0.22 &-0.12  & grad07grad11 &-0.19 &-0.30 &-0.15  \\
grad02grad07 &0.00 &-0.18 &-0.11  & grad07grad12 &-0.19 &-0.33 &-0.16  \\
grad02grad08 &0.00 &-0.15 &-0.09  & grad07grad13 &-0.19 &-0.37 &-0.17  \\
grad02grad09 &0.00 &-0.07 &-0.06  & grad07grad14 &-0.19 &-0.41 &-0.17  \\
grad02grad10 &0.00 &-0.04 &-0.03  & grad07grad15 &-0.19 &-0.44 &-0.18  \\
grad02grad11 &0.00 &-0.30 &-0.14  & grad08grad08 &-0.15 &-0.15 &-0.10  \\
grad02grad12 &0.00 &-0.33 &-0.15  & grad08grad09 &-0.15 &-0.07 &-0.08  \\
grad02grad13 &0.00 &-0.37 &-0.16  & grad08grad10 &-0.15 &-0.04 &-0.06  \\
grad02grad14 &0.00 &-0.41 &-0.17  & grad08grad11 &-0.15 &-0.30 &-0.14  \\
grad02grad15 &0.00 &-0.44 &-0.17  & grad08grad12 &-0.15 &-0.33 &-0.14  \\
grad03grad03 &-0.13 &-0.13 &-0.09  & grad08grad13 &-0.15 &-0.37 &-0.15  \\
grad03grad04 &-0.13 &-0.13 &-0.07  & grad08grad14 &-0.15 &-0.41 &-0.16  \\
grad03grad05 &-0.13 &0.00 &-0.04  & grad08grad15 &-0.15 &-0.44 &-0.16  \\
grad03grad06 &-0.13 &-0.22 &-0.11  & grad09grad09 &-0.07 &-0.07 &-0.05  \\
grad03grad07 &-0.13 &-0.18 &-0.11  & grad09grad10 &-0.07 &-0.04 &-0.04  \\
grad03grad08 &-0.13 &-0.15 &-0.09  & grad02grad11 &-0.00 &-0.30 &-0.14  \\
grad03grad09 &-0.13 &-0.07 &-0.07  & grad09grad12 &-0.07 &-0.33 &-0.11  \\
grad03grad10 &-0.13 &-0.04 &-0.06  & grad09grad13 &-0.07 &-0.37 &-0.11  \\
grad03grad11 &-0.13 &-0.30 &-0.13  & grad09grad14 &-0.07 &-0.41 &-0.12  \\
grad03grad12 &-0.13 &-0.33 &-0.14  & grad09grad15 &-0.07 &-0.44 &-0.12  \\
grad03grad13 &-0.13 &-0.37 &-0.14  & grad10grad10 &-0.04 &-0.04 &-0.03  \\
grad03grad14 &-0.13 &-0.41 &-0.15  & grad10grad11 &-0.04 &-0.30 &-0.08  \\
grad03grad15 &-0.13 &-0.44 &-0.15  & grad10grad12 &-0.04 &-0.33 &-0.09  \\
grad04grad04 &-0.13 &-0.13 &-0.09  & grad10grad13 &-0.04 &-0.37 &-0.09  \\
grad04grad05 &-0.13 &0.00 &-0.04  & grad10grad14 &-0.04 &-0.41 &-0.10  \\
grad04grad06 &-0.13 &-0.22 &-0.14  & grad10grad15 &-0.04 &-0.44 &-0.10  \\
grad04grad07 &-0.13 &-0.18 &-0.13  & grad11grad11 &-0.30 &-0.30 &-0.19  \\
grad04grad08 &-0.13 &-0.15 &-0.11  & grad11grad12 &-0.30 &-0.33 &-0.20  \\
grad04grad09 &-0.13 &-0.07 &-0.08  & grad11grad13 &-0.30 &-0.37 &-0.21  \\
grad04grad10 &-0.13 &-0.04 &-0.06  & grad11grad14 &-0.30 &-0.41 &-0.22  \\
grad04grad11 &-0.13 &-0.30 &-0.16  & grad11grad15 &-0.30 &-0.44 &-0.22  \\
grad04grad12 &-0.13 &-0.33 &-0.17  & grad12grad12 &-0.33 &-0.33 &-0.21  \\
grad04grad13 &-0.13 &-0.37 &-0.18  & grad12grad13 &-0.33 &-0.37 &-0.22  \\
grad04grad14 &-0.13 &-0.41 &-0.19  & grad12grad14 &-0.33 &-0.41 &-0.23  \\
grad04grad15 &-0.13 &-0.44 &-0.20  & grad12grad15 &-0.33 &-0.44 &-0.24  \\
grad05grad05 &0.00 &0.00 &0.00  & grad13grad13 &-0.37 &-0.37 &-0.23  \\
grad05grad06 &0.00 &-0.22 &-0.10  & grad13grad14 &-0.37 &-0.41 &-0.24  \\
grad05grad07 &0.00 &-0.18 &-0.09  & grad13grad15 &-0.37 &-0.44 &-0.25  \\
grad05grad08 &0.00 &-0.15 &-0.07  & grad14grad14 &-0.41 &-0.41 &-0.25  \\
grad05grad09 &0.00 &-0.07 &-0.05  & grad14grad15 &-0.41 &-0.44 &-0.26  \\
grad05grad10 &0.00 &-0.04 &-0.03  & grad15grad15 &-0.44 &-0.44 &-0.27  \\
\hline\hline
\end{tabular}
\end{table*}

\section{Conclusions}\label{concl}

By means of a set of N-body simulations involving equal mass elliptical
galaxies with no gas, we attempted to understand if dissipationless
mergers always cause the flattening of the initial galaxy gradient,
under what conditions does flattening happen, are there situations
where a gradient could become steeper, what is the amount of flattening
or steeping typically produced, and how does this evolution depend on
orbital and morphological properties of the galaxies in these interacting
systems.  Our purpose is to explore how the metallicity gradients in
dry mergers depends of the structure and metallicity gradients of the
galaxies involved in the merger.  Specifically, we would like to answer:
Could dry mergers lead to metallicity gradients of the ellipticals
observed in the local Universe?

From this modeling, we conclude that: \\
1. The ratio of the remnant and the
initial galaxy slopes span a wide range of values, up to values greater
than 1 (with values greater than one resulting only when companions have
gradients twice the progenitor).\\
2.  For a merger between two ellipticals
having identical initial metallicity slope (i.e., companion and galaxy
slopes being the same), the metallicity profile of the remnant flattens,
with a final gradient about 0.6 times the initial one. \\
3.  Ellipticals can
maintain their original pre-merger metallicity gradient if the companion
slope is sufficiently steep.\\
4.   The final remnant gradient does not depend
on the orbits of the progenitor ellipticals.  \\
5. While the final gradient does depend on the initial concentration of the progenitors, the amount of flattening (i.e. remnant slope/progenitor slope) does not. \\

Given the diversity in outcomes of the mergers, we conclude that dry
mergers do not violate any observational constraints on the systematic
characteristics of metallicity gradients in local ellipticals.  In fact,
dry mergers might also explain many of the characteristics of metallicity
gradients.  There are no systematic trends in magnitude of metallicity
gradients in ellipticals save one, metallicity gradients generally become
steeper with increasing mass, but with an increasing dispersion \citep{ogando05}.  This observation is
logically explained within the context of our simulations:  dry mergers lead to an overall decrease in the gradient, if the companion slope is not steep enough (companion slope/galaxy slope less than 2).
We note however, that as the metallicity gradient gets flatter, it
becomes easier for an elliptical to merge with another elliptical with
sufficiently steep slope to actually increase the final metallicity
gradient of the remnant. 
The reader should note that this mechanism
requires the existence at any given time of progenitors with enough steep slopes, perhaps formed
through a pure monolithic channel.  Thus, if dry merging is important, we would
expect the variance in the metallicity gradient to be the largest amongst
the most massive elliptical consistent which is what is observed.  Moreover,
since steepening can occur, especially for ellipticals with no or shallow
gradients, such a model also predicts that ellipticals will only rarely
show very shallow gradients.  This is something that has perhaps been
observed \citep[][]{ogando05,annibali07} and is therefore an argument in
favor of dry merging, not against it.

\section*{Acknowledgments} 
MDL wishes to thank the Centre National de la Recherche Scientifique
(CNRS) for its continuing support of his research.  PDM is supported
by a grant from the Agence Nationale de la Recherche (ANR) in France.
The authors wish to thank the referee, for his prompt and constructive report.

\begin{appendix} %
\section{Mixing processes and their dipendence on the orbital parameters }\label{app1}

\begin{figure}
   \centering
\includegraphics[width=3.cm,angle=270] {fig20.ps}
\includegraphics[width=3.cm,angle=270] {fig21.ps}
\caption{Evolution of the metallicity profiles in some major mergers involving two 2iE0 galaxies. Metallicity is shown versus the distance from the galaxy center for the two galaxies before the interaction (black and thick gray solid lines) and for the final remnant for different orbits (see legends). The values on the abscissa are shown in units of the half-mass radius $r_{50}$ of the corresponding galaxy.
\label{fig1app1}}
\end{figure}

\begin{figure}
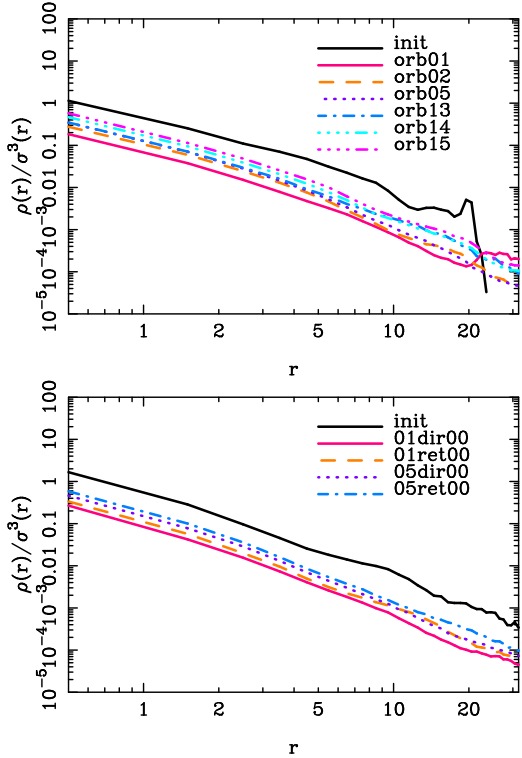

   \centering
\includegraphics[width=5cm,angle=270] {fig22.ps}
   \centering
\includegraphics[width=5cm,angle=270] {fig23.ps}
\caption{Evolution of the spherically averaged $Q(r)=\rho(r)/\sigma^3(r)$ during major dry mergers. Top panel: The  profile  $Q(r)$ is shown as a function of the distance $r$ (in kpc) from the galaxy center for the progenitor gE0 galaxy (solid thick black line) and for the remnant galaxies, for different orbits (see legend). Bottom panel: Same as the previous panel, but this time the merger involves two 2iE0 galaxies, whose initial $Q(r)$ is shown by a black solid line. Different orientation between the orbital angular momentum and the galaxy spins have been taken into account in this case (direct and retograde orbits), as well as different orbital energies and angular momenta (orbit 01 and 05).  \label{fig2app1}}
\end{figure}

\begin{figure}
   \centering
\includegraphics[width=5cm,angle=270] {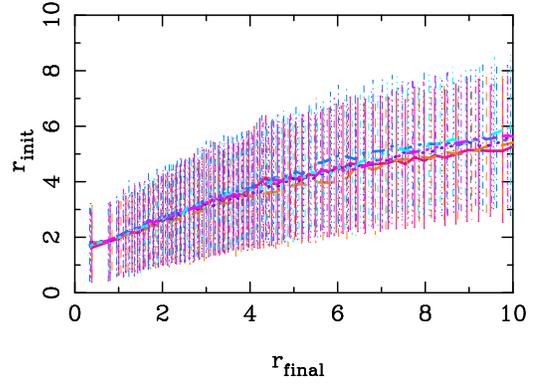}
\caption{Redistribution of stellar particles in the final remnants of gE0-gE0 mergers. For every stellar particle in the remnant, we have evaluated its position in the remnant (through its distance $r_{final}$ from the remnant center) and the position it had in the progenitor galaxy (through its distance $r_{init}$ from the progenitor center). The averaged quantities are then plotted, each curve corresponding to a different orbit, as explained in the legend of Fig.\ref{fig2app1}, left panel. The corresponding standard deviations are also shown. Note that particles are redistributed in a very similar way in the merger remnants (similar averages and dispersions), independent of the orbit of the encounter.
\label{fig3app1}}
\end{figure}

In Sect.\ref{fiducial} we have seen that the evolution of the metallicity profile for the fiducial merger between two gE0 galaxies does not depend on the energy and angular momentum of the orbit.
As discussed in Sect.~\ref{model}, the gE0 models represent quite idealized initial conditions, being spherical systems, without rotation. The final product of a merger of two not rotating spherical galaxies is typically a triaxial system, with some amount of rotation, due to the redistribution of the orbital angular momentum in internal one. Thus it can be interesting to understand if  "hierarchical" initial conditions (the initial galaxies being the remnants of previous mergers) still produce final metallicity profiles which show similar properties, independently on the parameters of the interaction.  A different alignement between the initial galaxy spins and the orbital angular momentum can lead, for example, to different tidal effects in the interacting pairs. Can this also lead to final metallicity profiles depending on the initial orbital energy and angular momentum? In other words, are the results found in Sect.\ref{fiducial} confirmed when repeated hierarchical mergers are considered?\\
To assess the generality of the results,  we first  simulated the merger of two iE0 galaxies (not rotating spherical galaxies whose masses are equal to half of the gE0 mass) and then two identical copies of the corresponding remnants were merged again, changing the initial orbital energy and angular momentum and its orientation with respect to the galaxy spins.\\
The result of these experiments are shown in Fig.\ref{fig1app1}, where the metallicity profiles of 2iE0-2iE0 mergers are shown, for different orbits, corresponding to different initial values of the orbital angular momentum (orbit id=01 and 05) and to different alignments of this orbital angular momentum with respect to the galaxy spins (dir00 and ret00 corresponding to the most extreme cases, when the orbital angular momentum is respectively parallel or antiparallel to the galaxy spins). In all the cases examined, it results that the final metallicity profiles are independent on the orbit of the encounter, at least inside the half mass radius of the remnant galaxy. It is interesting to note that this conclusion, i.e. \emph{the independence of the remnant metallicity profiles on the orbit of the encounter}, does not depend on the morphological and kinematical properties of the initial systems either. Indeed it is valid when the merger involves two spherical symmetric, not rotating systems (Sect.\ref{fiducial}), as well as when it involves two triaxial, rotating galaxies (Fig.\ref{fig1app1}).\\

But where does this behaviour come from? In other words, how do mixing and phase-space density evolution proceed in collisionless major mergers and how do they depend on the orbital parameters of the encounter?
The collisionless Boltzmann equation tells us that the evolution of a collisionless system is characterized by the conservation of the six-dimensional phase-space density (also known as the "fine-grained" phase-space density).
The computation of this phase-space density being quite difficult, one usually computes the "coarse-grained" phase-space density, which represents the average of the fine-grained one over some finite volume in the phase-space, and which obeys one of the Mixing Theorems \citep[see for example][]{trem86} or the spherically averaged quantity $Q(r)=\rho(r)/\sigma^3(r)$ - being $\rho(r)$ the spherically averaged volume density and $\sigma(r)$ the velocity dispersion-  which still has the dimensions of a phase-space density, but being only a proxy of the coarse-grained distribution function, it does not satisfy any of the Mixing Theorems.\\
\citet{vass08} have recently studied the evolution of the phase-space density distribution of dark matter halos during equal-mass mergers. They have taken into account a variety of shapes for the dark matter halos (from steep cusps to core-like profiles), of different orbital parameters, as well as recursive mergers of dark halos, in order to study the effect of a hierarchical building up of halos on the evolution of the phase-space density profiles. They studied both the evolution of the coarse-grained distribution function $F(r)$ and of its spherically averaged proxy $Q(r)$, showing that they evolve in a similar way during the merging process, independently on the orbital properties of the encounter and on the number of repeated collisions. In all cases, the inner slopes and the overall shapes of the phase-space density distribution of the merger remnant are close to that of the progenitor system.\\ We too found similar results  evaluating the $Q(r)$ profile of our remnant galaxies, for different orbital initial conditions. As shown in Fig.\ref{fig2app1}, mergers of spherically symmetric  (upper panel) and of triaxial (lower panel) systems lead to an evolution of the spherically averaged $Q(r)$ which is independent on the orbit and which retains memory of the progenitor profile.\\
This independency in the evolution of the phase-space distribution on the orbital properties of the encounter leads also to very similar evolutions in the spatial distribution of particles after the collisions. In other words, stellar particles redistribute themselves in the remnant galaxy in a very similar way, independently on the orbits (see Fig.\ref{fig3app1}), at least inside the remnant half-mass radius ($r_{50}\simeq 7$kpc for gE0-gE0 mergers). This ultimately gives rise to the very similar metallicity profiles found in Fig.~\ref{zprofile} and  Fig.~\ref{fig1app1}.

\end{appendix}

\end{document}